*sensors*

*Article*

# Latency Reduction in Vehicular Sensing Applications by Dynamic 5G User Plane Function Allocation with Session Continuity

Pablo Fondo-Ferreiro [†], David Candal-Ventureira [†], Francisco Javier González-Castaño and Felipe Gil-Castiñeira *

Information Technologies Group, atlanTTic, University of Vigo, 36310 Vigo, Spain; pfondo@gti.uvigo.es (P.F.-F.); dcandal@gti.uvigo.es (D.C.-V.); javier@gti.uvigo.es (F.J.G.-C.)
* Correspondence: xil@gti.uvigo.es; Tel.: +34-986-818665
† These authors contributed equally to this work.

**Abstract:** Vehicle automation is driving the integration of advanced sensors and new applications that demand high-quality information, such as collaborative sensing for enhanced situational awareness. In this work, we considered a vehicular sensing scenario supported by 5G communications, in which vehicle sensor data need to be sent to edge computing resources with stringent latency constraints. To ensure low latency with the resources available, we propose an optimization framework that deploys User Plane Functions (UPFs) dynamically at the edge to minimize the number of network hops between the vehicles and them. The proposed framework relies on a practical Software-Defined-Networking (SDN)-based mechanism that allows seamless re-assignment of vehicles to UPFs while maintaining session and service continuity. We propose and evaluate different UPF allocation algorithms that reduce communications latency compared to static, random, and centralized deployment baselines. Our results demonstrated that the dynamic allocation of UPFs can support latency-critical applications that would be unfeasible otherwise.

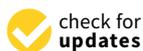

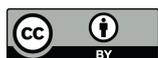

## 1. Introduction

Vehicles have been equipped during recent years with a wide range of sensors such as odometers, Global Positioning System (GPS) receivers, distance sensors, cameras, etc. Furthermore, the new autonomous vehicles require more and better sensors for identifying and localizing objects in the surroundings of the vehicle. By combining sensors and communication capabilities, vehicles become excellent platforms that enable a wide range of applications including environmental surveillance and traffic monitoring [1]. Thus, vehicular sensing has been tackled from many different perspectives. Environmental sensors have been installed in many different types of vehicles, both sensing-oriented and in opportunistic sensing scenarios. In some cases, engine or navigational sensors and data transceivers have been used for secondary purposes, instead of installing on-board ad hoc sensors. In fact, cell phones enable use cases of great interest, because they may serve as on-board gateways between engine Controller Area Network (CAN) buses and the Internet; they can contribute data from their own sensors (magnetometers, camera, accelerometers, etc.) [2], and they can provide measurements of cellular network coverage, to cite some possibilities. Indeed, vehicular sensing is a relevant subfield of vehicular networks. For example, in [2], the authors reviewed the way sensor information is collected, stored, and harvested using intervehicular communications, as well as their supporting infrastructure (e.g., centralized and distributed storage in the wired Internet). More recently, in [3], collaborative sensing amongst vehicles for automated driving, where full coverage must be guaranteed in risky spots, was studied, by analyzing the role of Roadside Units (RSUs) in assisting vehicular sensing.

5G networks [4,5] are a key enabler contributing to the enrichment of the subfield. They provide existing powerful smartphones connections with unprecedented performance in terms of access bandwidth, latency and scalability and are supported by intelligent packet cores and fully softwarized/virtualized networks, which span across radio access, backhaul, and core resources [6–8]. They also include distributed network-side computing resources at the fog, edge, and cloud layers [9,10].

However, all this flexibility comes at the cost of powerful dynamic management paradigms, which, in some cases, must be assisted by artificial intelligence. From the point of view of vehicular sensing, a crucial requirement is to take vehicular data as soon as possible to the network-side computing resources that process these data to make decisions or to provide additional information to drivers, vehicles, or the infrastructure.

Consequently, in this paper, we are interested in a solution to take vehicular data to the User Plane Function (UPF) modules of a 5G network core as efficiently as possible since, and from that moment on, those data will be readily accessible to the computing resources at the network edge. For this purpose, we relied on our previous Software-Defined-Networking (SDN)-based solution in [11]. A differential benefit of this solution is that, no matter how many UPF allocations are necessary, the data sessions are not interrupted.

By combining 5G gNBs with the associated computing resources and our solution to place anchor points at optimal locations, we can adapt the network to the requirements of the UEs in terms of latency. This enables applications such as collaborative sensing for enhanced situational awareness, where vehicles send large amounts of Light Detection and Ranging (LiDAR) data and other high-resolution sensor information to applications that fuse them together to build a high-precision view of the environment in real time. This information is then distributed to all the vehicles in the area so that they can avoid obstacles or make decisions [12–14]. All these procedures (collection of information, processing, and distribution of the resulting information) have to be completed under strict time constraints (as low as 10 ms [15]). Besides, many other applications, such as Augmented Reality (AR) for providing rich information to drivers, should also be supported by powerful onboard computers or remote servers deployed at the edge [16]. In the second case, even a small latency of 50 ms can cause an error of 50 cm in the information displayed [17]. In order to reach such latency values, it is necessary to reduce the number of hops between the vehicles and the edge computing infrastructure, as each hop may increase the latency in the range from hundreds of microseconds to a few milliseconds for fiber optic and radio links (depending on the technology [18]). Moreover, the network processing delay in each hop is not negligible [19].

The main contributions of this paper are:

- The proposal of latency-reduction algorithms for the dynamic allocation of UPFs in 5G vehicular sensing scenarios;
- The evaluation of the proposed algorithms in terms of end-user latency and execution time using a publicly available dataset with vehicular mobility traces and Base Station (BS) deployments;
- The comparison of the proposed algorithms with baseline allocation strategies.

The rest of this paper is organized as follows: Section 2 discusses the background of this work. Section 3 describes the problem we address. Section 4 explains our proposed solution. Section 5 describes the methodology, and Section 6 presents our results. Finally, Section 7 summarizes our findings and concludes the paper.

## 2. Background and Related Work

Vehicular sensor data have been exploited in many applications and research studies. Classic examples of sensing-oriented vehicles are space balloons, environmental satellites, and oceanographic vessels, to name just a few. Even though these platforms have been operated for decades, all of them are still useful for research [20–22]. New specialized sensing vehicles, such as drones, keep appearing [23,24], raising their own novel inter-networking problems [25]. Opportunistic vehicular sensing, on the other hand, relies on the currently highly dense transport routes, such as city streets, air lanes, and motorways. Therefore, by installing low-cost sensors in public transportation vehicles, for instance, the operation costs can be minimized. This idea is not new: the National Aeronautics and Space Administration (NASA) deployed sensors in 65 commercial aircraft in 2006 to create the Tropospheric Airborne Meteorological Data Reporting (TAMDAR) moisture data repository for weather forecasting [26,27]. In [28–31], we and other authors independently proposed public-transportation-based systems for pollution and road condition evaluation. Later, the research in [32] formalized this paradigm by defining and measuring the coverage depending on the traffic routes, by determining the relationship between the coverage quality and the number of vehicles and selecting the minimum number of vehicles to achieve a target coverage quality.

The advent of cell phones, as high-performance computing devices linked to the Internet through broadband data networks, has revolutionized vehicular sensing. Cell phones can contribute their GPS positions to monitor urban dynamics [33]. Road conditions can be estimated from cell phone accelerometer data [34]. Much of the required preprocessing can be executed at user terminals themselves at the cost of some user incentives [35].

Before the advent of 3G mobile communications in the late 1990s, extracting large volumes of data from mobile sensors was unfeasible. Indeed, mobile sensors were data mules. As a result, solutions such as Delay-Tolerant Networking (DTN) were applied to take their data to the Internet when connection opportunities arose [31,36] (these solutions are currently still used in extreme environments [37]). This was a logical and straightforward extension of the paradigm of Intermittently Connected Delay-Tolerant Wireless Sensor Networks (ICDT-WSNs), a branch of WSNs that was surveyed in [38].

As previously said, the field has been revolutionized by 5G networks [4,5]. In vehicular 5G communications, a thriving research field by itself [39], transferring large sensing data volumes will no longer be an issue, at least in urban environments in the short term, provided that the density of base stations is high enough [40,41]. Of course, new challenges will appear when the massive IoT becomes a reality [42,43] due to scaling issues, but 5G communications have been conceived of to deal with these from the onset [44]. A more immediate challenge currently is taking sensor data to edge computing resources, where decisions are made (for example, for crossroad safety applications [45]), at minimum latency. We are interested in studying 5G solutions for this purpose in vehicular sensing. In this regard, we assumed that the computing resources for sensing applications are scalable and, at least for the preprocessing tasks, stateless, and thus that they are available at all base stations. That is, we are not interested in redeploying computing resources following the movement of the vehicles, another interesting problem that was tackled in [46] for instance, but in taking sensor data to those resources as efficiently as possible.

The deployment of UPFs at the edge has been recently explored as a promising strategy to support novel applications [47–49]. In [47], the authors elaborated on a neural network-based model for predicting traffic load to perform a proactive autoscaling of UPFs, as well as on Service Function Chaining (SFC) placement. Then, they formulated the problem of allocating the required number of UPFs to the desired locations as an Integer Linear Program (ILP). However, the focus of their work was the prediction of the traffic load, and they did not consider the possibility of increasing the number of UPFs in order to further reduce the latency. Other works, such as [48], have studied a joint user association, Virtual Network Function (VNF) placement, and resource allocation with a Mixed-Integer Linear Programming (MILP) formulation. A similar model was used in [49] for placing end-to-end slices modeled as SFCs. In the latter, the authors approximated the solution to optimize the number of VNF migrations and the network utilization with a heuristic algorithm that satisfies the requirements of the slices.

None of these works considered the challenge of seamlessly redirecting the traffic of the UEs from the base station to the desired UPF. Besides, in this work, we studied the trade-off between the number of UPFs deployed in the network and the latency perceived by the users in vehicular sensing scenarios.

For the purposes of the seamless dynamic reconfiguration of user communications between different locations, in our previous work in [11], we proposed an SDN-based mechanism that changes the serving anchor point for end-users (i.e., UPF in 5G networks) between different edge locations while ensuring session and service continuity. The mechanism consists of a dynamic deployment of anchor points in the desired edge locations through Network Function Virtualization (NFV) and the synchronization of the context information of the anchor points between the different locations through the SDN. Finally, SDN technology was also used to reconfigure the network to redirect the traffic flows of each User Equipment (UE) to the desired anchor point. To the best of our knowledge, UPF allocation subject to session continuity (as supported by a practical solution) has not been previously considered.

## 3. Problem Statement

In this paper, we considered a 5G network scenario composed of multiple base stations (gNBs) distributed across a 2D surface. The base stations provide wireless coverage to UE-enabled vehicles that move across the surface. As the vehicles move, their association with the base stations may change following their movement. The vehicles capture information from the environment through different onboard sensors and send this information to instances of a stateless distributed server-side application. The communications between the vehicles and the instances take place according to 3GPP standards: a General Packet Radio Service (GPRS) Tunneling Protocol (GTP) tunnel is established between the gNB and a UPF anchor point for each UE. The UPF is the termination endpoint for the communications of each UE, whose data are then forwarded to the server application. For the sake of simplicity, we considered a 1-1 mapping between base stations and Multi-access Edge Computing (MEC) hosts, which correspond to the nodes of a simple graph. The edges of the graph represent direct communication channels between the base stations (e.g., fiber links). Nevertheless, we remark that the collocation of the MEC hosts with the BSs is just an example of a deployment strategy for simplifying the problem statement. The proposed algorithms in this work can be applied to other deployment scenarios [10] (e.g., collocation of the MEC host with transmission nodes, collocation of the MEC host with a network aggregation point).

The role of the MEC hosts in our scenario is two-fold: they are not only used to host the lightweight distributed stateless server-side applications that receive the sensor data from the vehicles, but also to deploy UPF instances.

We considered a time-slotted model whose network intelligence periodically determines in which nodes UPFs are deployed to minimize the latency perceived by the UEs for the next slot based on the distribution of UEs in the current slot. We measured the latency perceived by each UE as the minimum number of hops between the base station the UE is attached to and a base station containing a UPF. In this context, we refer to a base station that contains users attached to it during the current time slot as an active base station.

Changing the serving UPF for a given UE is challenging because it may cause session and service disruption. This is related to the change of the endpoint of the GTP tunnel to a different UPF. To ensure session and service continuity, it is necessary to replicate the state of the previous UPF to the new one and properly reconfigure the network before the reassignment takes place. To this end, we propose to use the SDN-based mechanism defined in [11], which allows seamlessly changing the serving UPF for a UE. The proposed solution combines the benefits of SSC Modes 1 and 3 defined in 3GPP TS 23.501 [50] (i.e., changing the IP anchor point and also maintaining the IP address allocated to the UE). As described in Section IV.C of [11], the mechanism follows a make-before-break approach to ensure that the new UPF is completely ready and operational before actually diverting to it the traffic of the UEs.

In a practical implementation, our solution in [11] can be used both to dynamically deploy the UPFs at the desired edge locations and also to divert the traffic of each UE to the closest UPF.

## 4. Proposed Solution

We wish to show that the dynamic intelligent allocation of UPFs in the network edge combined with a lightweight network reconfiguration can be used to reduce the overall latency since the data are captured by the vehicle sensors until they are processed by the network-side application. The elapsed time for applying the desired UPF allocation in each time slot must be short compared to the length of the intervals where their decisions are applied (i.e., the duration of the time slots). This includes both the elapsed time for executing the UPF allocation algorithms and also the times required for deploying the UPFs at the target locations and reconfiguring the underlying SDN network for steering the traffic of each user to the closest UPF. Using our proposed mechanism in [11], the times required for deploying a lightweight UPF and reconfiguring the SDN network are about 900 ms and 40 ms, respectively [51]. This previous result establishes a lower bound of 1 s on the duration of the time slot. That is, the time elapsed between two consecutive executions of the dynamic UPF allocation algorithm must exceed 1 s to guarantee system stability. In practice, values above 5 s are recommended in order to minimize the overhead by too frequent network reconfigurations.

Figure 1 shows a simplified view of the proposed architecture. In detail, with respect to the standard 5G architecture, our proposal introduces SDN switches at the edge sites (i.e., close to the gNBs), which are configured to forward the traffic of each user to the desired UPF. The SDN switches are controlled by an SDN controller. The decisions on UPFs locations are made by the UPF allocation algorithm. The decision is communicated to the NFV Management and Orchestration (MANO) platform of the operator, which triggers the instantiation of the UPF VNFs at the desired MEC hosts. Then, the SDN controller is notified to reconfigure the network and divert the traffic of each UE to the closest UPF, by leveraging our mechanism in [11].

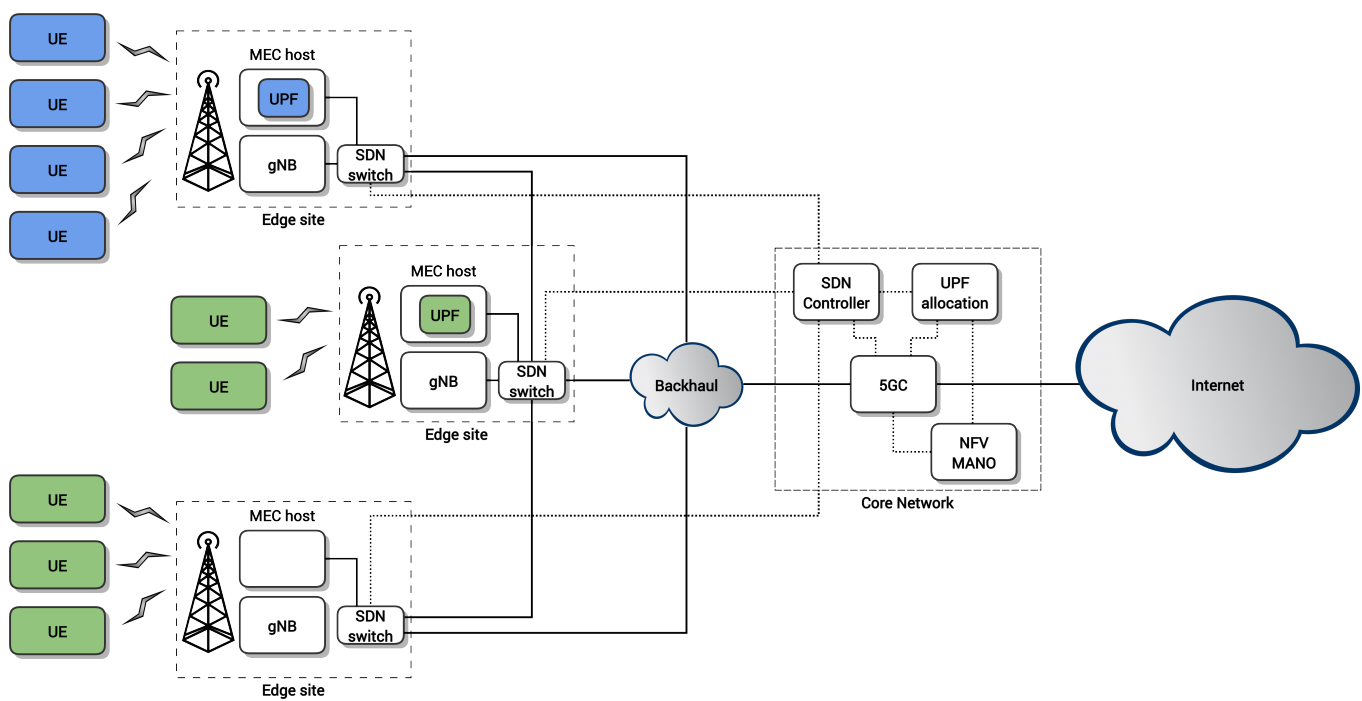

**Figure 1.** Architecture of the proposed solution.

The inputs required by the UPF allocation algorithm are as follows:

- The network graph is composed of the set of base stations as nodes and the corresponding links between them;
- The set of active users for each base station.

Then, the output generated by the algorithm is the set of nodes where the UPFs must be deployed. With this in mind, we considered the following UPF allocation algorithms:

- Static K-means: A K-means clustering of all the base stations in the network is calculated. A UPF is deployed at the closest base station to the location of each cluster center based on Euclidean distances. This strategy is static, given the locations of the base stations, and it does not depend on the user distribution. Therefore, it does not involve any dynamic redeployment of the UPFs;
- Random: A trivial approach that picks the base stations where the UPFs are deployed at random, according to a discrete uniform distribution. Note that the scenario in which just one UPF is used is equivalent to a centralized UPF deployment;
- Greedy percentile: This iteratively picks the node that reduces the 90th-percentile worst-case latency perceived by the UEs. In the case of multiple nodes introducing the same 90th-percentile latency, the tie is broken in favor of the node with the largest reduction of the average latency perceived by the UEs;
- Greedy average: This iteratively picks the base station with the largest reduction of the average latency perceived by the UEs;
- K-means: A K-means clustering of the active base stations looking for as many clusters as UPFs we are interested in deploying. A UPF is deployed at the closest base station to the location of each cluster center based on Euclidean distances. This is analogous to the proposal in [52];
- K-means greedy average: A K-means clustering of the active base stations is calculated. A UPF is deployed in each cluster by determining the node with the largest reduction of the average latency perceived by the UEs attached to base stations in the cluster;
- Louvain modularity greedy average: Nodes are clustered according to the Louvain modularity maximization [53]. A UPF is deployed in each cluster by selecting the node with the largest reduction of the average latency perceived by the UEs attached to base stations in the cluster.

The static K-means and random algorithms were used as baselines. The greedy percentile algorithm seeks to reduce the 90th-percentile worst-case latency perceived by the UEs. The greedy average is a simplified version of the latter to reduce its computational complexity. The K-means algorithm follows a state-of-the-art approach described in [52] to determine the location of the UPFs as cluster centers, to reduce the computational complexity as well. We also propose a variant of this approach called K-means greedy average, which greedily determines where a UPF is deployed inside each cluster. The objective of this variant is to further reduce the latency perceived. Finally, the Louvain modularity greedy average algorithm relies on graph theory techniques for identifying communities [53] inside the network topology. Similar to the K-means greedy average, this last strategy selects in a greedy manner the placement of each UPF inside each community. Other techniques from graph theory (e.g., Girvan–Newman community detection [54]) were also considered in this work, but we did not include them in the final results because they did not provide improvements with respect to the Louvain modularity greedy average.

Since the goal of our experiments was to analyze the trade-off between the number of deployed UPFs and the latency achieved, we set the number of clusters of the K-means and Louvain modularity algorithms to the same fixed value as the number of UPFs being evaluated.

As a final remark, in the case of using a different MEC deployment strategy (i.e., UPFs restricted to specific base stations), the algorithms can be directly generalized by considering its constraints.

## 5. Methodology

In this section, we describe the methodology we followed in this work to evaluate the proposed UPF allocation algorithms in a vehicular sensing scenario.

We first describe the datasets we used, the processing we applied, and some insights based on an exploratory analysis of their information. Then, we elaborate on the metrics we used to assess the performance of the proposed algorithms through simulations.

*5.1. Dataset*

To evaluate our proposal in a realistic scenario, we used two publicly available datasets with the locations of real cellular base stations and realistic vehicular traffic traces in an urban environment [55].

The dataset with mobility traces is available at http://kolntrace.project.citi-lab.fr/koln.tr.bz2 (accessed on 17 November 2021). It corresponds to a 400 m$^2$ area in the city of Cologne, Germany. The dataset containing the base stations was obtained from public German databases in 2012, and it is available at http://kolntrace.project.citi-lab.fr/koln_bs-deployment-D1_fixed.log (accessed on 17 November 2021). It contains the Cartesian coordinates of 247 base stations. We assumed that each base station is connected to other base stations in their vicinity, conforming a single connected graph, and that some base stations are equipped with a local computing infrastructure on which the operator can deploy low-latency services for its users based on the MEC paradigm.

To build the single connected graph, we determined the links between the base stations as follows:

- First, each base station is connected to all the base stations in its surroundings less than $DISTANCE\_LINK\_THRESHOLD$ away. This results in a set of connected components;
- Then, we built the single connected graph by iteratively finding the pair of nodes from the largest and second-largest components at the shortest distance from each other and set an edge connecting these two nodes. This way, at each iteration, the connected components are joined to the largest component.

In our simulations, we set $DISTANCE\_LINK\_THRESHOLD$ to 500 m. The first part of the procedure resulted in 177 connected components. At the end of the procedure, the resulting graph had 293 undirected edges. Figure 2 shows the topology of the resulting interconnection network. Figure 3 shows the histogram of the minimum distances between pairs of nodes. As we can observe, they are normally distributed with an average of around 18 hops between every pair of base stations.

The dataset of vehicular mobility traces consists of almost 20 GB of data of a period of 24 h. In total, there are over 354 million entries, corresponding to more than 700 thousand individual trips. Each entry contains the Cartesian coordinates of the vehicle location at each time instant plus the speed of the vehicle.

The UEs in this dataset do not produce measurements at all timestamps. We considered that UEs are off during the periods in which they have no data. To assign a base station to each UE, we used the following procedure, also used in related works (e.g., [46]):

- If the UE has just been enabled (that is, there was no information for the UE in the immediately preceding time slot), we assign the UE to the base station with the minimum path loss. To compute the path loss, we used the expression and parameters for a non-line-of-sight urban scenario of [56]. For the sake of clarity, the formula is reproduced next:

$$\overline{PL}(d)[dB] = \alpha + 10 \cdot \beta \cdot log(d) + X_\sigma, \qquad (1)$$

where $\alpha$ and $\beta$ are the least-squares fits of the floating intercept and the slope, respectively, $d$ is the distance between the base station and the UE, and $X_\sigma$ is a Gaussian random variable with zero mean and standard deviation $\sigma$ that represents the shadowing effect;

- Otherwise, we evaluated the path loss of the nearest base station and the one to which the UE was attached in the previous round. If the path loss to the nearest base station is lower than that to the base station to which the UE was associated plus an additional hysteresis threshold $\epsilon$, the UE roams to this base station. Else, the UE remains associated with the previous base station. In our simulations, we set $\epsilon$ to 2 dB, as recommended in [57].

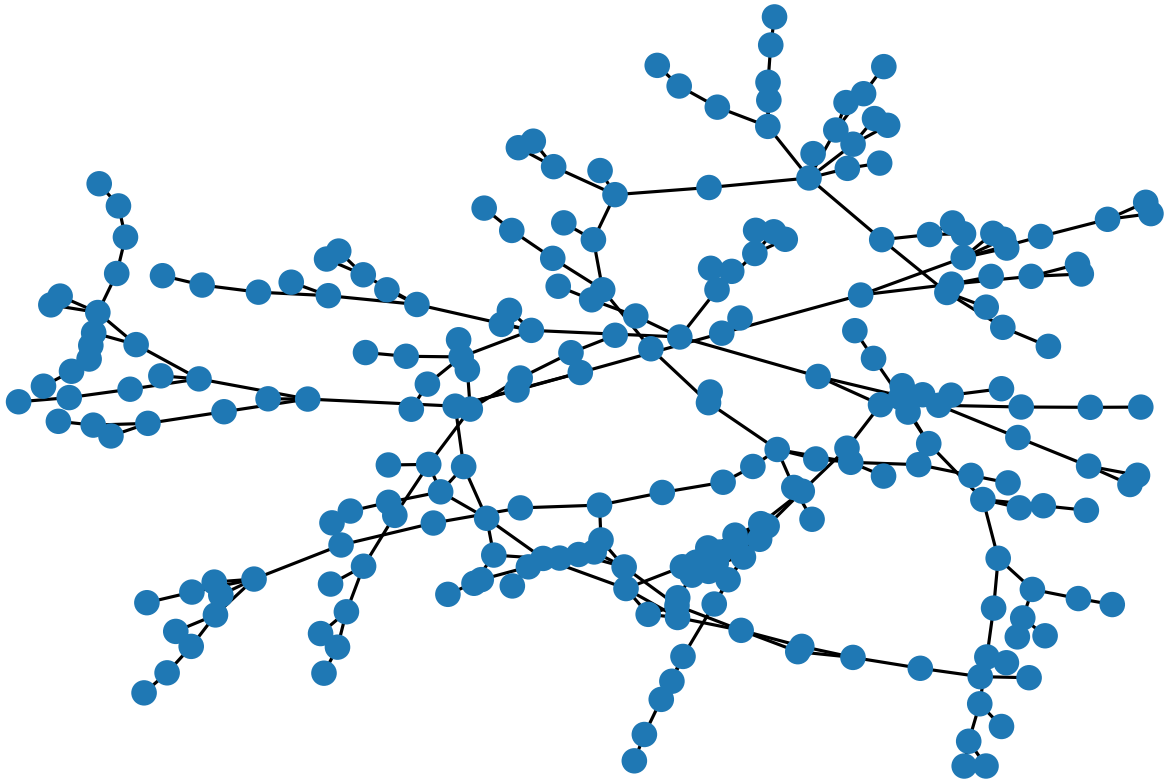

**Figure 2.** Resulting interconnection network topology.

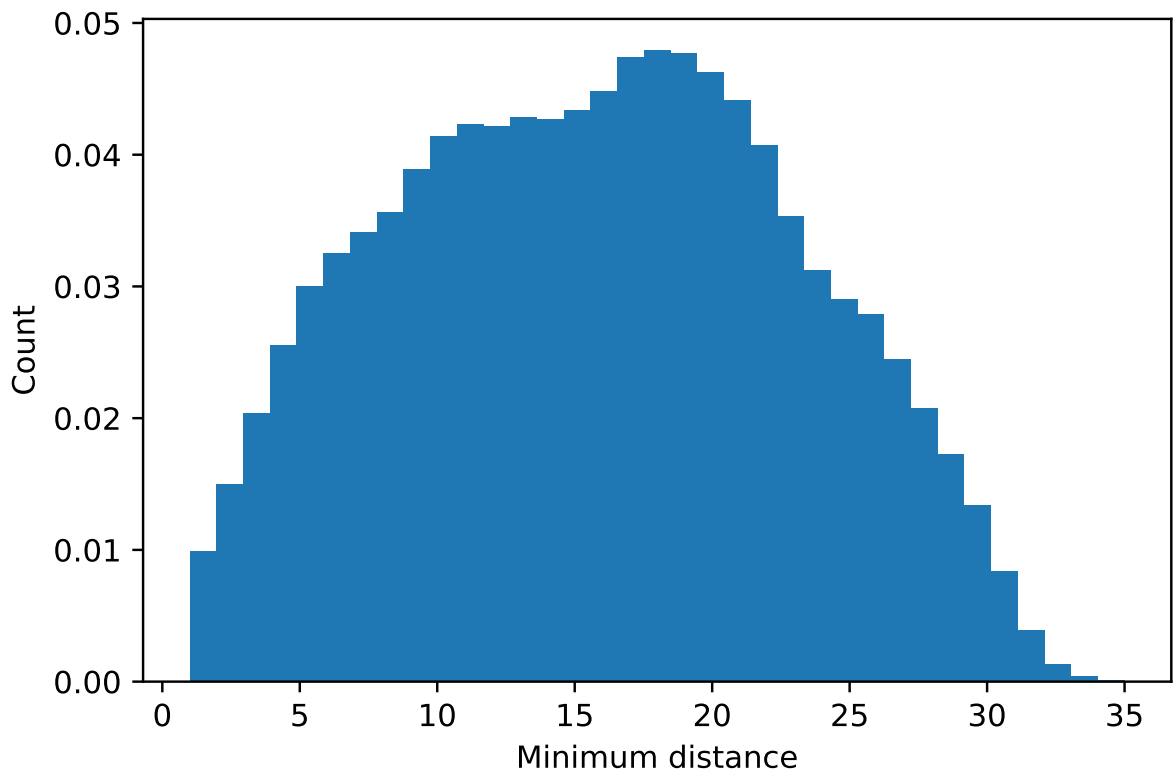

**Figure 3.** Histogram of the minimum distances (number of hops) between every pair of base stations in the resulting graph.

*5.2. Evaluation Metrics*

We evaluated the proposed algorithms with two metrics, *UE latency* and *execution time*:

- UE latency is a measure of the latency perceived by each UE since it transmits the sensed information until it is received at the server-side application. In this work, we characterized it as the minimum number of hops from the base station to which the UE is connected to a node where a UPF is deployed. As an aggregate measure of all individual UE latencies, we considered the 90th-percentile of the latency;
- Execution time is the time the UPF allocation algorithm takes to determine in which nodes the UPFs will be deployed. We calculated this as the elapsed time since the UPF allocation module receives the input data until the output is generated by the corresponding algorithm. That is, this time reflects only the decision time at each time slot.

## 6. Results and Discussion

For the reproducibility of our results, the Python code used for the simulations is available in a public repository [58] under an open-source license. The simulations were executed using the PyPy Python implementation [59] on top of commodity hardware (Intel Core i9-9900K CPU @ 3.60GHz Intel, Santa Clara, CA, USA).

In our simulations, we set the duration of the time slot to 5 s. Figure 4 shows the 90th-percentile of the latency perceived by the UEs when the percentage of BSs containing a UPF increases. The results are provided with 95% confidence intervals. First, we can observe that, by increasing the percentage of BSs that contain a UPF, the latency decreased in general, as expected (there was an exception in the K-means clustering algorithms that will be explained in detail later). The higher latencies correspond to the random algorithm, rendering it useless. The other algorithms succeeded in reducing the number of hops to the UPF by more than 30% compared to that random baseline. In detail, the lowest latency values were attained by the greedy percentile and greedy average algorithms, which provided very similar results. The K-means greedy average algorithm was the next best one followed by the Louvain modularity greedy average, which performed slightly worse for percentages of BSs with UPFs lower than 5%. Among the dynamic allocation algorithms, the basic K-means algorithm introduced the highest latency. Finally, the latency values achieved by the static K-means algorithm were very similar to those achieved by the dynamic version of the K-means algorithm, except for slight fluctuations depicted at certain percentages of BSs with UPFs. Both K-means and static K-means exhibited a strange behavior for very low percentages, where the latency grew when increasing the number of UPFs. This was due to the behavior of the algorithm, which selects the nodes for allocating UPFs based on the centers of the clusters identified by the K-means clustering algorithm without considering the actual distribution of UEs.

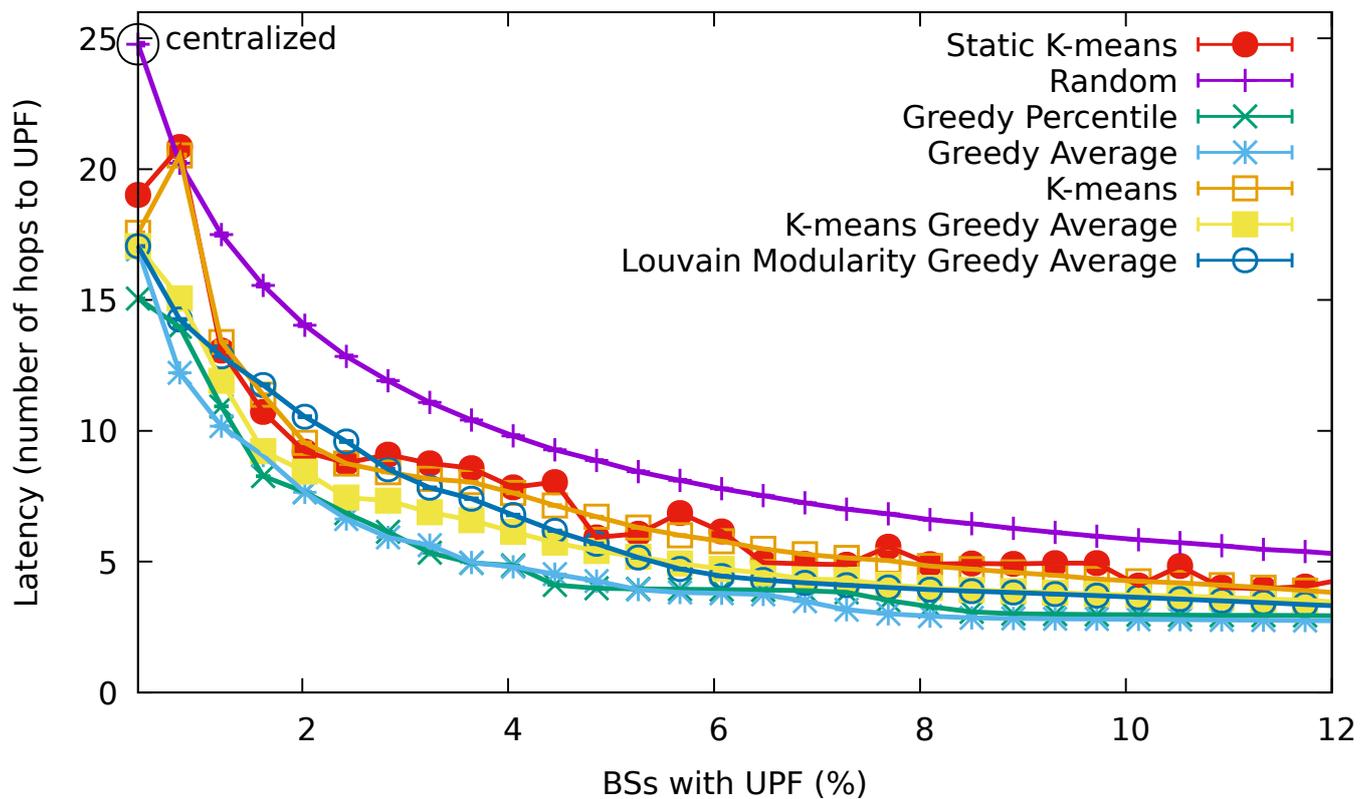

**Figure 4.** The 90th-percentile of the latency (number of hops) perceived by the UEs for a growing percentage of base stations containing a UPF.

To understand the unexpected behavior in the results of the K-means algorithm, we decided to take a closer look at its UPF allocations. Specifically, we studied the allocation at a time slot of the simulation with a representative number of UEs.

Figures 5–7 show the distribution of the 4112 UEs in the 3000th time slot of the simulation for allocating 1, 2, and 3 UPFs using the K-means algorithm, respectively. The obtained results for the 90th-percentile latency were 17 hops, 20 hops, and 13 hops, respectively. As we can observe, most of the UEs were located near the (15, 15) point, which turned out to be the center of the cluster when considering just one UPF (see Figure 5), and therefore, the latency perceived by the UEs was in line with the other greedy-based algorithms. However, when two clusters were considered, none of the centers were placed in the area with the highest density of UEs (see Figure 6), which resulted in a performance degradation with respect to allocating just one UPF. Indeed, the resulting latency value was similar to that obtained by the random algorithm for two UPFs. In the case of three UPFs, we can observe that one of the clusters was located in the area with the highest density, while the other two covered the areas with a lower density of UEs (see Figure 7), further reducing the latency perceived to similar values obtained by the other greedy-based algorithms.

Finally, note that a simplistic traditional deployment consisting of a single centralized UPF in the core network serving all the base stations would introduce a higher latency than any of the proposed algorithms using multiple edge UPFs.

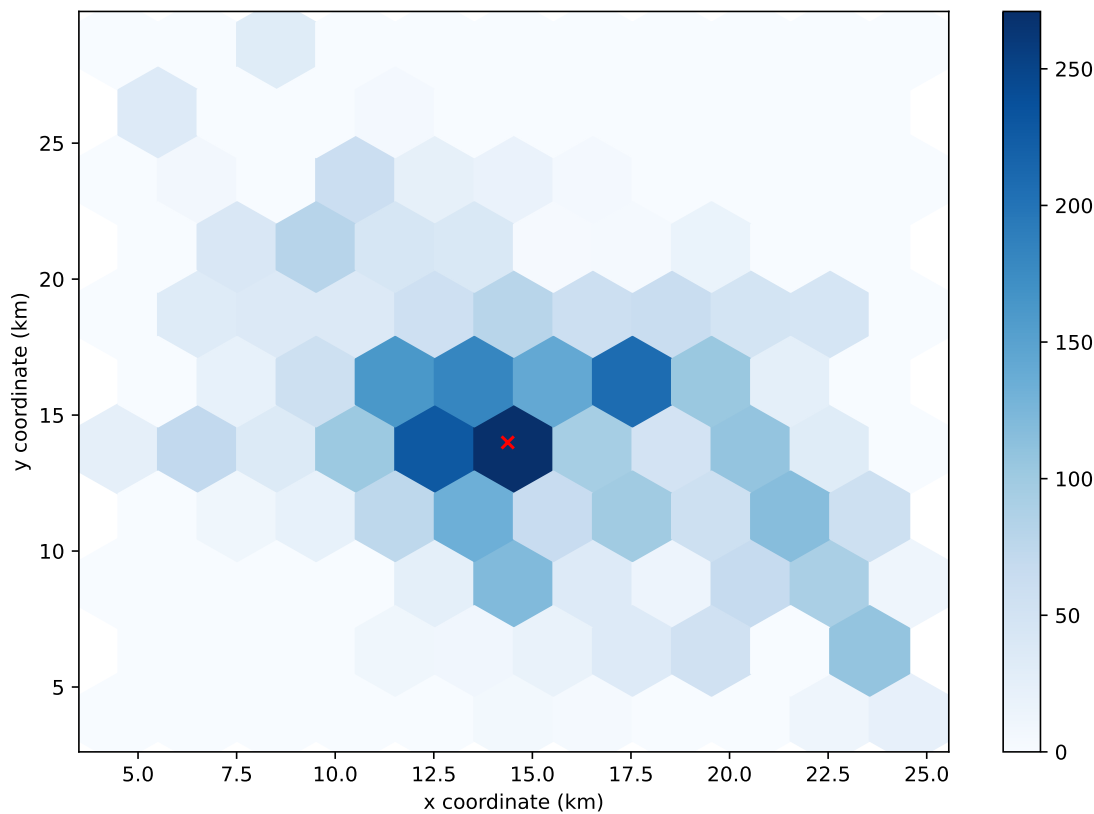

**Figure 5.** Distribution of 4112 UEs in the 3000th time slot of the simulation. The red cross marks the location where 1 UPF is allocated using the K-means algorithm.

Figure 8 shows the average execution time for the different algorithms as the number of base stations containing a UPF increases. As we can observe, the greedy percentile algorithm is computationally expensive because it must evaluate the 90th-percentile latency that would be perceived by the UEs for every possible UPF deployment at each iteration, which results in a computational complexity of $\mathcal{O}(B^4 \cdot U)$, where $B$ is the number of base stations and $U$ the percentage of BSs with UPF. This complexity noticeably reduces in the case of the greedy average algorithm, which only evaluates the average latency that would be perceived with each of the potential UPF placements, for a computational complexity of $\mathcal{O}(B^3 \cdot U)$. Interestingly, the latency values achieved are strongly similar to those achieved by the greedy percentile algorithm. Finally, the K-means algorithm is able to further reduce this time by executing a lightweight clustering algorithm and then calculating the closest base stations to the cluster centers. However, determining the UPF based on the location of the centers is not the most efficient approach, because a naive implementation requires checking all the base stations against each cluster center. In addition, the results obtained by this algorithm were very sensitive to outliers, and besides, they can be improved by greedily picking the node for deploying a UPF inside each cluster, as shown by the K-means greedy average algorithm. Finally, the Louvain modularity greedy average algorithm improves the efficiency by considering a lightweight hierarchical clustering algorithm, which, in addition, exploits the graph structure of the nodes in the network with a clustering of the nodes based on a graph modularity maximization. The performance gains in terms of execution time were substantial compared with the K-means algorithms, for only a minor degradation in the latency value achieved in the case of low percentages of base stations with UPFs.

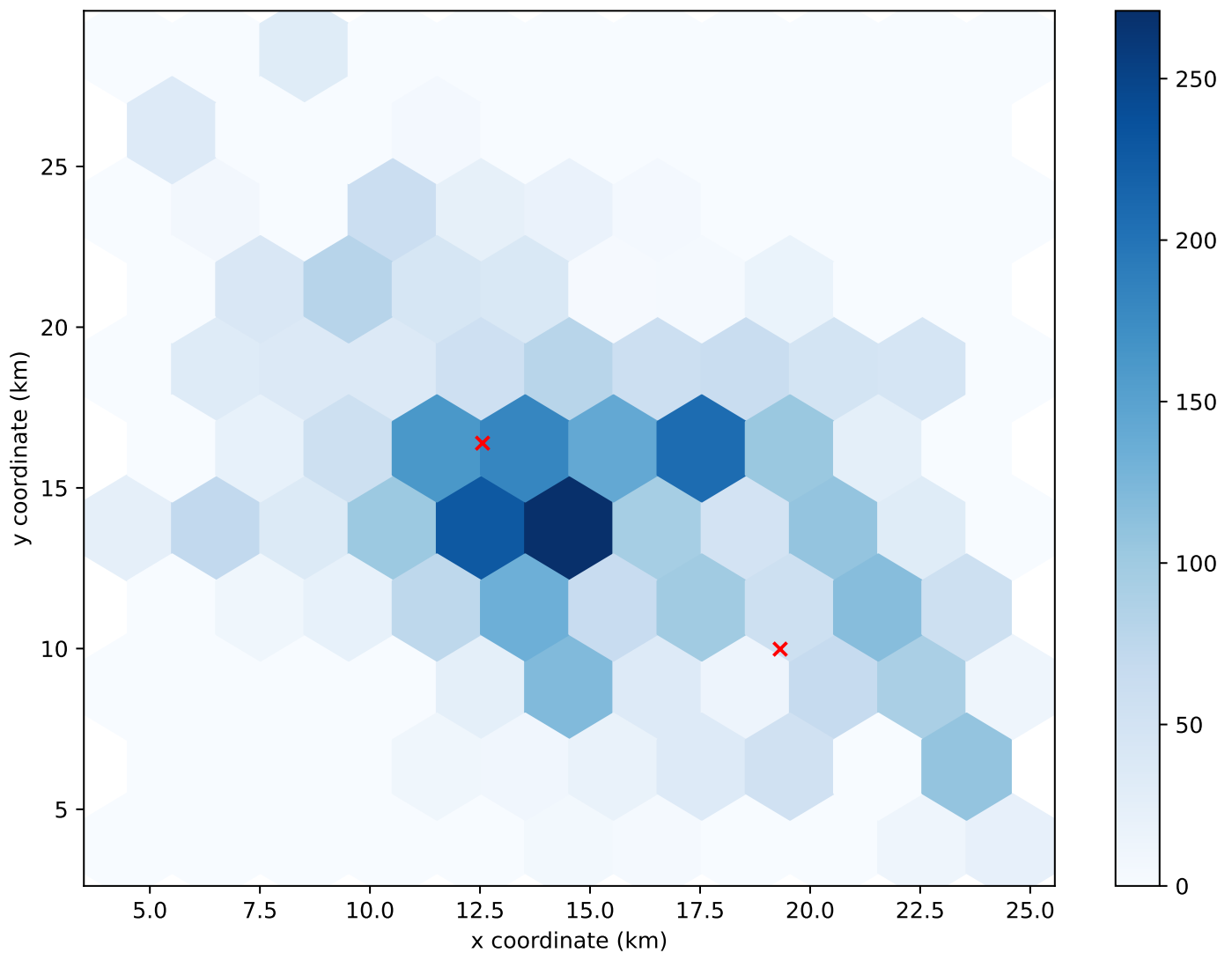

**Figure 6.** Distribution of 4112 UEs in the 3000th time slot of the simulation. The red crosses mark the location where 2 UPFs are allocated using the K-means algorithm.

In a practical situation, the actual number of required UPFs can be determined from the curves in Figure 4 according to the latency requirements of the application. For example, considering the stringent latency requirements of 5 ms for vehicular communications [60] and assuming an optimistic latency estimate of 1 ms per hop, our results showed that a centralized deployment will not be able to satisfy the requirements (as this deployment yielded a latency of about 20 ms). Baseline strategies such as random deployments and static K-means would require deploying UPFs in 10% and 7% of the base stations, respectively. Dynamic optimization strategies such as the Louvain modularity greedy average that we propose achieve a 5% reduction of these values in a very short execution time. Finally, pure greedy strategies can further reduce them by a 3%, yet at the cost of increased computational complexity.

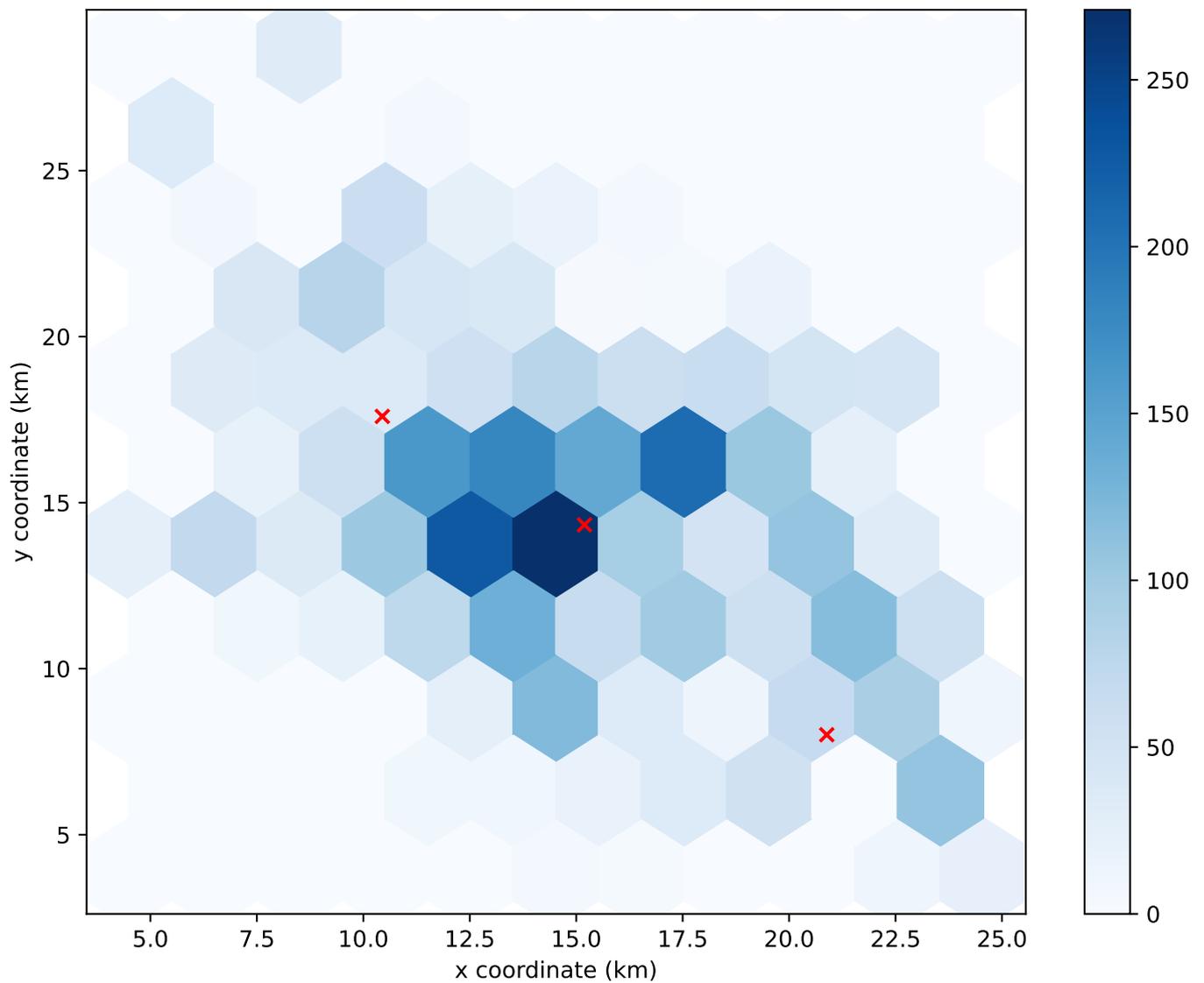

**Figure 7.** Distribution of 4112 UEs in the 3000th time slot of the simulation. The red crosses mark the location where 3 UPFs are allocated using the K-means algorithm.

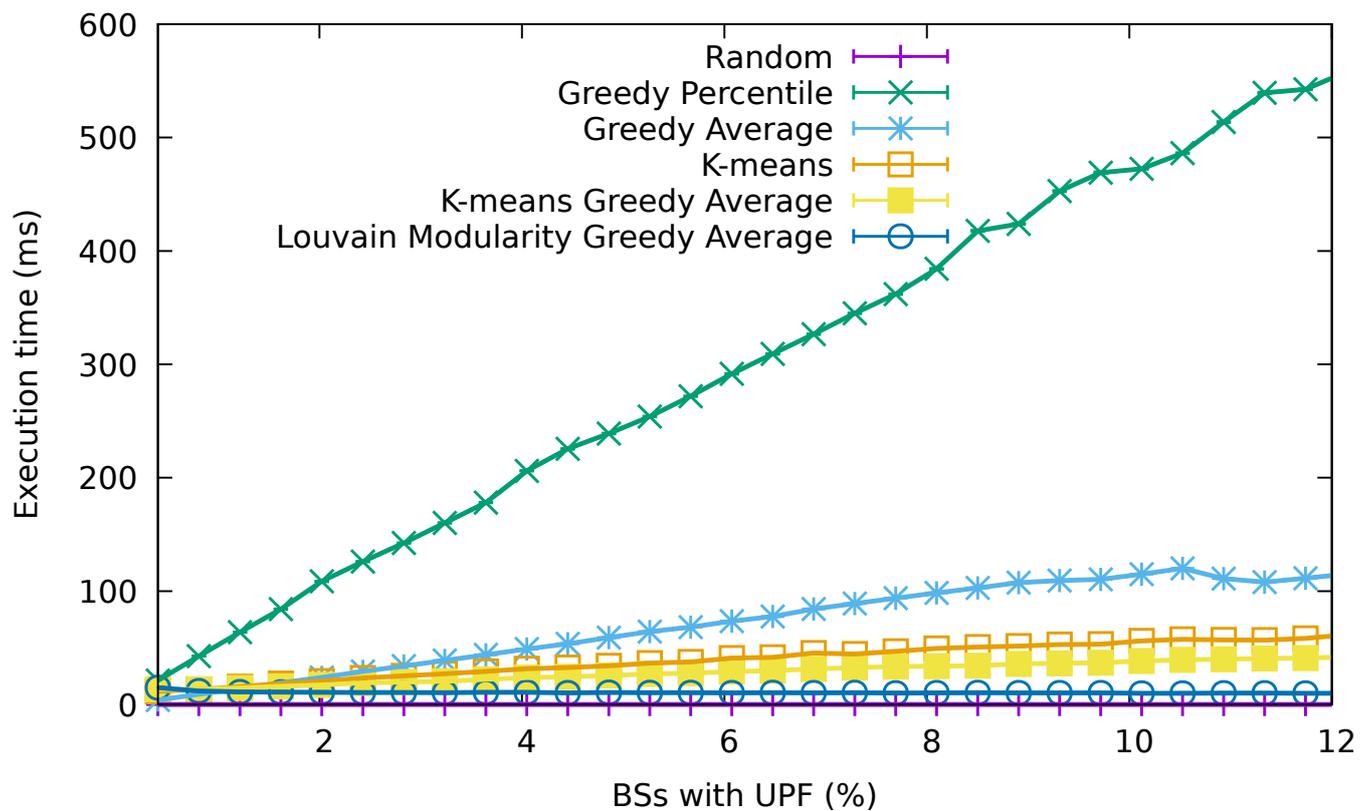

**Figure 8.** Average execution time for the different algorithms for a growing percentage of base stations containing a UPF.

## 7. Conclusions

The increasing amount and quality of the information provided by the sensors integrated into vehicles can be used to provide advanced services for drivers, autonomous vehicles, smart cities, etc. The complexity of the generated information requires a large amount of computing power that may not be possible to install inside the vehicle, but thanks to the quality of the communication channels, it is viable to perform the processing of the information in remote servers, following the cloud or edge paradigms. Nevertheless, highly demanding services such as collaborative enhanced situational awareness require fusing and analyzing the collected information and redistributing the results to the participating nodes. The latency between the vehicles and the UPFs should be minimized to take vehicular data to edge computing resources and provide results on time for the vehicle to understand the surrounding environment. In this work, we proposed a dynamic lightweight optimization framework that achieves this goal by continuously optimizing the deployment of UPF anchor points for each UE. The proposed framework is compatible with our SDN-based mechanism described in [11], which guarantees session continuity even during UPF reallocation.

We compared the performance of different UPF allocation algorithms in terms of end-user latency achieved and execution time required. Our results revealed that centralized and static deployments cannot satisfy the requirements of latency-sensitive applications. However, our proposal based on the Louvain modularity greedy algorithm provides a good trade-off between the two previous metrics, allowing the deployment of applications such as enhanced situational awareness for low percentages of BSs with UPFs.

**Author Contributions:** Conceptualization, F.J.G.-C.; methodology, F.J.G.-C.; software, P.F.-F. and D.C.-V.; validation, D.C.-V., P.F.-F. and F.G.-C.; formal analysis, F.J.G.-C.; investigation, P.F.-F. and D.C.-V.; resources, F.G.-C.; data curation, P.F.-F. and D.C.-V.; writing—original draft preparation, P.F.-F. and F.J.G.-C.; writing—review and editing, F.G.-C. and D.C.-V.; visualization, D.C.-V.; supervision, F.G.-C. and F.J.G.-C.; project administration, F.G.-C. and F.J.G.-C.; funding acquisition, F.G.-C., F.J.G.-C. and P.F.-F. All authors have read and agreed to the published version of the manuscript.

**Funding:** This research was partially funded by a "la Caixa" Foundation (ID 100010434) fellowship (LCF/BQ/ES18/11670020), Ministerio de Ciencia e Innovación Grant PID2020-116329GB-C21 and Xunta de Galicia Grants GRC2018/05, ED341D-R2016/012, and IN854A 2020/01 (Factory competitiveness and electromobility through innovation, FACENDO 4.0).

**Data Availability Statement:** The data analyzed in this study can be directly generated from the code available in the public repository [58] using as input the data provided in the following publicly available datasets: vehicle mobility traces http://kolntrace.project.citi-lab.fr/koln.tr.bz2 (accessed on 17 November 2021) and base station deployments http://kolntrace.project.citi-lab.fr/koln_bs-deployment-D1_fixed.log (accessed on 17 November 2021).

**Conflicts of Interest:** The authors declare no conflict of interest. The funders had no role in the design of the study; in the collection, analyses, or interpretation of the data; in the writing of the manuscript; nor in the decision to publish the results.

## Abbreviations

The following abbreviations are used in this manuscript:

| | |
|---|---|
| SDN | Software-Defined Networking |
| NFV | Network Function Virtualization |
| VNF | Virtual Network Function |
| MANO | Management and Orchestration |
| MEC | Multi-access Edge Computing |
| UE | User Equipment |
| UPF | User Plane Function |
| GPRS | General Packet Radio Service |
| GTP | GPRS Tunneling Protocol |
| DTN | Delay-Tolerant Networking |
| ICDT-WSN | Intermittently Connected Delay-Tolerant Wireless Sensor Network |
| WSN | Wireless Sensor Network |
| CAN | Controller Area Network |
| RSU | Roadside Unit |
| AR | Augmented Reality |
| TAMDAR | Tropospheric Airborne Meteorological Data Reporting |
| GPS | Global Positioning System |
| NASA | National Aeronautics and Space Administration |
| LiDAR | Light Detection and Ranging |
| SFC | Service Function Chaining |
| ILP | Integer Linear Programming |
| MILP | Mixed-Integer Linear Programming |
| BS | Base Station |